\documentclass{ptapap}
\usepackage{natbib}

\author{Geza Kovacs}[KON]
\affil[KON]{Konkoly Observatory of the Hungarian Academy of Sciences\\
  Konkoly Thege M. \'ut 13-15, Budapest, Hungary}

\title{Big Signals, Small Signals, Systematics}

\begin{document}

\maketitle

\begin{abstract}
We examine the role of instrumental systematics in the search for small amplitude signal components in the presence of large amplitude signals. Current analyses of variable stars dealing with the above situation seem to ignore the effects of systematics, albeit the consideration of systematics is quite routine in the field of extrasolar planets. We show that simultaneous filtering of the large amplitude component and the systematics leads to residuals with much better potentials to detect any hidden small amplitude components. The method is illustrated by selected examples from the {\it Kepler} database. 
\end{abstract}

%
%
\section{Introduction}

With continuing data acquisition and developing instrumentation, the search for small amplitude signals in astronomical time series remains an important focus point in variable star astronomy \citep[see, e.g., ][on the early detection of the $0.01$\thinspace mag amplitude variation of the $\delta$~Scuti star $\epsilon$~Cep]{breger1966}. Until the photometric space missions {\it CoRoT} and {\it Kepler}, ground-based facilities usually provided time series with a precision of $1$\%--$0.1$\% per data point. Although this could have already been sufficient to access the sub-millimagnitude regime, except for the data gathered by coordinated observations (e.g., WET\footnote{Whole Earth Telescope, http://www.physics.udel.edu/gp/darc/wet/}  and DSN\footnote{Delta Scuti Network, https://www.univie.ac.at/tops/dsn/intro.html}), or, by long- time monitoring (e.g., OGLE\footnote{Optical Gravitational Lensing Experiment, http://ogle.astrouw.edu.pl/}), this goal was rarely attainable. The situation has somewhat changed with the advent of wide field surveys in search for extrasolar transiting planets (e.g., WASP\footnote{Wide Angle Search for Planets, \\
North: http://www.superwasp.org/, \\
South: http://www.superwasp.org/waspsouth.htm}, \\
HAT\footnote{Hungarian-made Automated Telescope, \\ 
North: https://hatnet.org/, \\ 
South: https://hatsouth.org/}, \\ 
KELT\footnote{Kilodegree Extremely Little Telescope, \\
North: http://www.astronomy.ohio-state.edu/keltnorth/Telescope.html, \\
South: https://my.vanderbilt.edu/keltsouth/}). Nevertheless, these surveys still remained in the regime of precision of a few mmag per data point \citep[but, because of the increased data volume, allowing to detect sub-mmag stellar variability in a large number of objects -- see][]{holdsworth2017}. 

The quantum leap that has led to the long-waited photometric discovery of solar-type oscillations, both in giants and in main sequence stars \citep{ridder2009, chaplin2011}, would have not been possible without the space missions. At the same time, large amplitude variables also attract a great deal of interest, since, in principle, they could also be the target of a high level of scrutiny similar to those showing solar-type pulsations or planetary transits. Indeed, as it was shown by the first discovery of the small amplitude modes in the large amplitude $\delta$~Scuti star AI~Vel by \cite{walraven1992}, small amplitude pulsations may occur in classical pulsators. Several other studies on large amplitude pulsators, made on space- and ground-based data, have led to the discovery of this type of small amplitude variabilities \citep[e.g.,][]{gruberbauer2007, moskalik2009}.

\begin{figure}[h]
\centering
  \includegraphics[width=.8\textwidth]{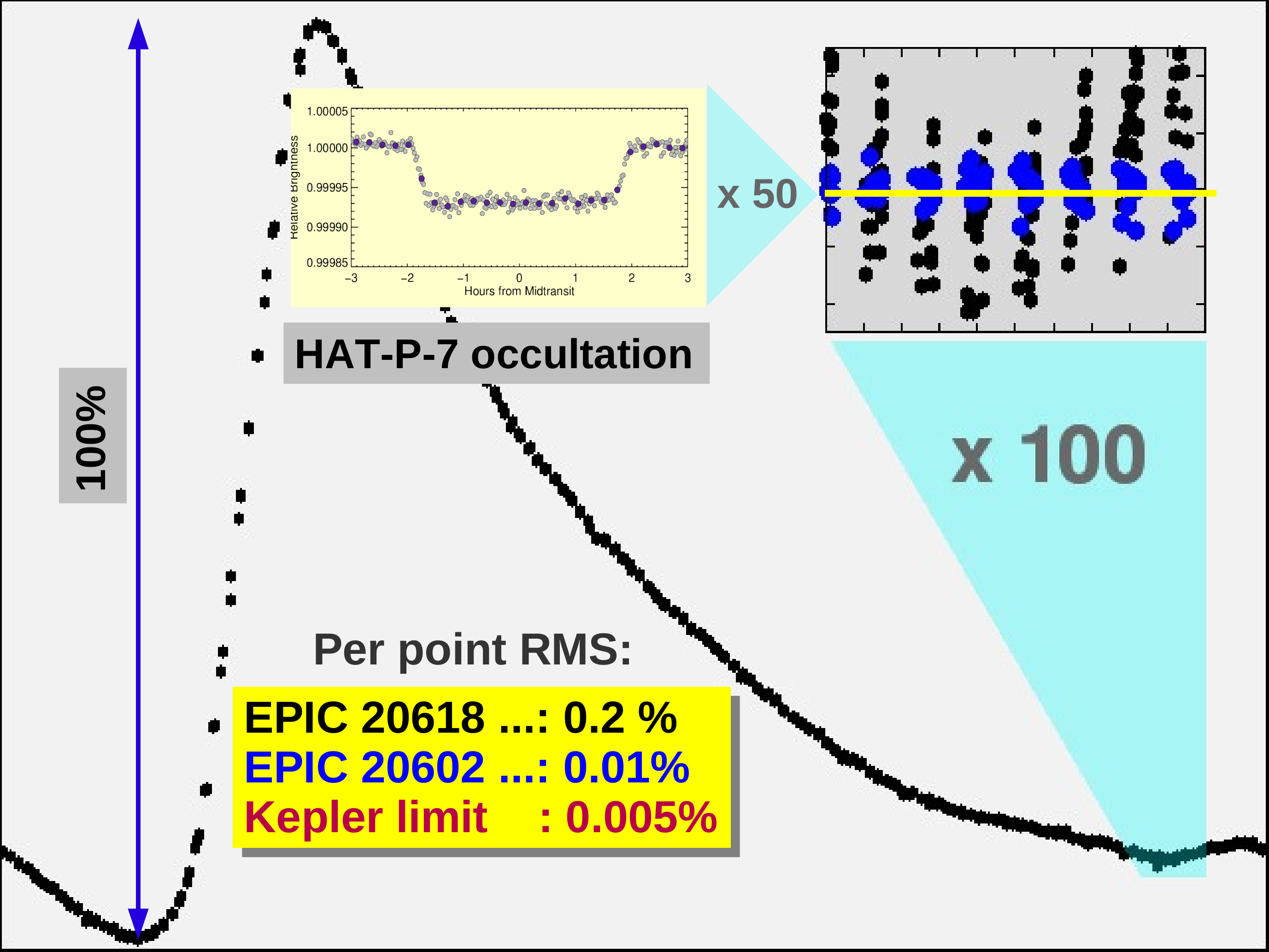}
\caption{Illustration of the range of variability covered by the {\it Kepler} space observatory from RR~Lyrae stars to the secondary eclipse of extrasolar planets. We plot the PDC-SAP folded light curve of a monoperiodic RR~Lyrae star EPIC\,206185368 \citep[black dots, see][]{armstrong2016} and that of a non-variable, EPIC\,206027866 (blue dots, folded with the same period and zoomed). {\it Total ranges:} inset (gray panel) 1\%; inset (pale yellow panel) $0.02$\%.}
\label{rr2pl}
\end{figure}

In Fig.~\ref{rr2pl},\footnote{The inset on HAT-P-7 in Fig.~\ref{rr2pl} has been reproduced from the tutorial of Andrew Vanderburg at https://www.cfa.harvard.edu/$\sim$avanderb/tutorial/tutorial2.html} we illustrate the enormity of the precision of the {\it Kepler} data with respect to the size of the light variation of a monoperiodic RR~Lyrae star in the {\it K2}/C03 field. The figure also shows that a `constant' star of the same brightness from the same field has a much smaller scatter. The purpose of this contribution is to highlight this difference and investigate possible sources of this discrepancy.

%
%

%
%
\section{Systematics and Large-Amplitude Signals}

When the range of variability spans through six orders of magnitude and the quality of the data allows us to investigate the full range, small effects, usually not considered under `normal' circumstances, may lead to incorrect signal representation and loss of components.

\begin{figure}[h]
\centering
\begin{minipage}{0.48\textwidth}
\includegraphics[width=\textwidth]{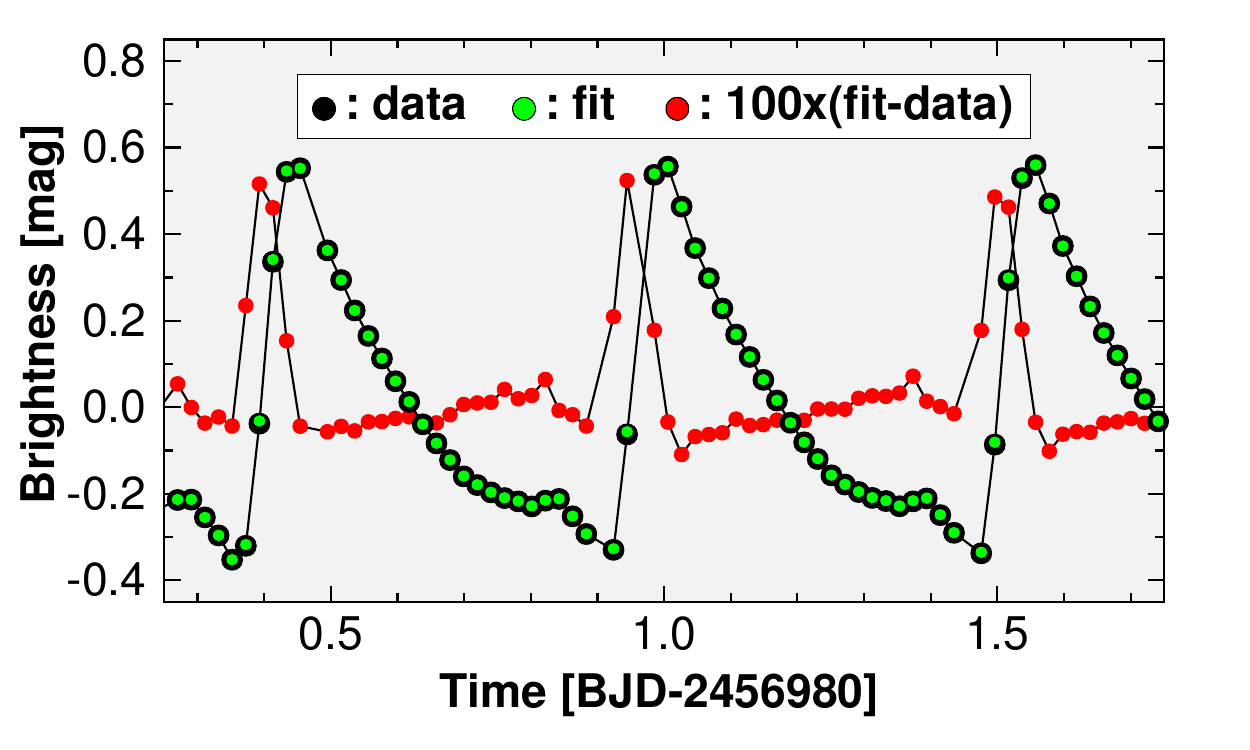}
\caption{A simple example on the importance of using precise frequencies in filtering out large amplitude signal components in search for small amplitude signal constituents. Artificial data are used on the {\it K2}/C03 time base as described in the text.}
\label{wrong-freq}
\end{minipage}
\quad
\begin{minipage}{0.48\textwidth}
\includegraphics[width=\textwidth]{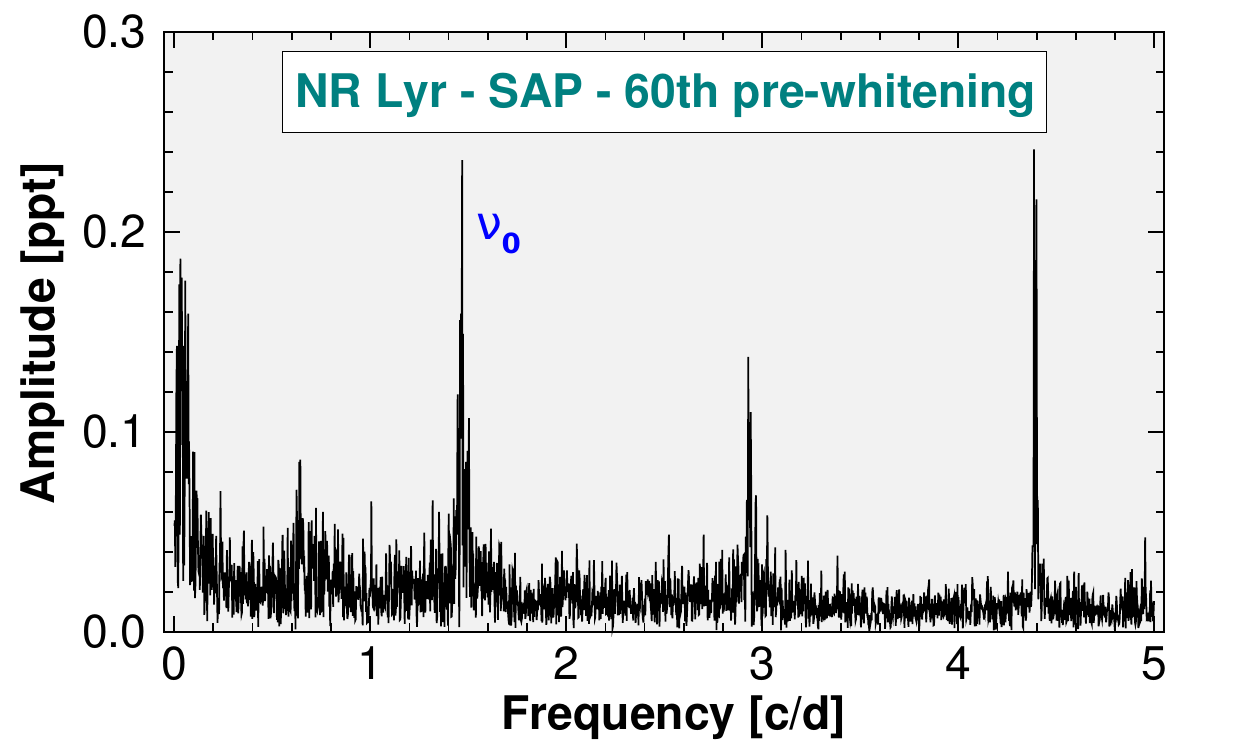}
\caption{An example of the power remaining at the pulsation frequency $\nu_0$ and its harmonics after many cycles of pre-whitenings of the large amplitude RR~Lyrae variable NR~Lyr from the {\it Kepler} field \citep[see][]{nemec2011,benko2018}.}
\label{spec-NRLyr}
\end{minipage}
\end{figure}

The simplest case is the error introduced by applying an imprecise pre-whitening frequency. We generated a time series on the timestamps of EPIC\,206027866 from {\it K2}/C03 by using the Fourier decomposition of the light curve of the well-known RR~Lyrae star RR~Cet. We added a Gaussian noise of $0.01$\% corresponding to a RMS of $\sim30$\thinspace ppm on a $6$\thinspace h time base, in a broad agreement with the current precision of the {\it K2} data \citep[e.g.,][]{vanderburg2014}. The frequency of the input signal was set equal to $1.8083$\thinspace c/d, whereas the trial frequency was purposely larger by $0.000015$\thinspace c/d (corresponding to a period difference of $0.0000046$\thinspace d). By performing a full Fourier fit (i.e., using the same number of harmonics we employed in the signal generation), we got the result shown in Fig.~\ref{wrong-freq}. In spite of the tiny phase shift of $0.001$ through the time span of $67$\thinspace days, the residual is much larger than the noise level and (obviously) biased toward the steeply varying part of the light curve. The improper fit results in a long chain of pre-whitening cycles and may lead to the misinterpretation of the data.     

%
%

The need for repeated pre-whitenings of frequencies near the known frequency components of the large-amplitude variables is a quite general attribute of studies using space-based observations \citep[see the note on the $414$ frequency fit by ][in the analysis of the triple-mode RR~Lyrae star EPIC\,201585823]{kurtz2016}. In addition to the white noise component, as a result of instrumental systematics, we also have red noise. On the top of these there might be a real period change of unknown size on unknown time scales. Fig.~\ref{spec-NRLyr} shows the severity of this issue on the example of a supposedly monoperiodic RR~Lyrae star from the original {\it Kepler} field. We employed the simple aperture photometry (SAP) as implemented by \cite{benko2018}. Since the data did not pass through the {\it Kepler} Pre-search Data Conditioning \citep[PDC, see][]{smith2012} pipeline, one suspects instrumental effects as a possible source of the remaining power. This noise source interacts with the true signal and leaves traces of itself via the well-known convolution theorem: 
%
%
\begin{equation}
\label{eq:1}
\mathcal{F}[Y] = \mathcal{F}[S]*\mathcal{F}[Y_0]\,,
\end{equation}    
where $\mathcal{F}$ denotes the Fourier transform, $Y$ stands for the observed, $Y_0$ for the true signal, and $S$ denotes the instrumental signal response (e.g., pixel sensitivity) function.   

One of the possible reasons why stellar variability studies have the tendency of avoiding data passed through some post-processing aimed for cleaning the signal from systematics, is that the methods cause various degrees of degradation in the original signal. This can, in principle, be alleviated by making the signal search assuming a complete model \citep{Foreman-Mackey2015, aigrain2015, angus2016}, but it is computationally far more intensive and leads to somewhat lower detection efficiency than methods following the ``clean first then analyze'' rule \citep{kovacs2016b}. However, the situation is quite different in the presence of large amplitude signals. The periods of these signals are usually quite accurately known (at least to a first approximation) and can be comfortably filtered out from the time series, before getting into the more delicate analysis of the low amplitude components.

Briefly, using the framework of the Trend Filtering Algorithm \citep[TFA,][]{kovacs2005}, we first fit the data by satisfying the Least Squares condition for $\mathcal D$, built up from the residuals between the data \{$Y$\} and the signal model \{$F$\}, containing both the co-trending set \{$X$\} and the Fourier components \{$U$\} of the known large amplitude signal: 
%
%
\begin{eqnarray}
\label{eq:1}
\mathcal D & = & \sum_{i=1}^N \left[Y(i) - F(i)\right]^2 \,,\\
      F(i) & = & \sum_{j=1}^{N_{\rm TFA}}c_{j} X_{j}(i) + 
           \sum_{k=1}^{2N_{\rm FOUR}}d_{k} U_{k}(i)\,.
\end{eqnarray}    
With the \{$c$\}, \{$d$\} coefficients determined from the minimization of $\mathcal D$, we can pre-whiten the data with this first estimate of \{$Y$\} and search for new signals in the residuals. Please note that this method will also introduce some degradation of the unknown signal, because (as it is being unknown) it is not included in the model. Nevertheless, as many successful applications have shown over the past, this is still a quite efficient way of searching for faint signals primarily affected by systematics.

%
%
\section{An Example on the Method at Work}

We tested several stars from the {\it K2}/C03\footnote{We used the data processed by Self Flat Fielding \citep[SFF, see ][]{vanderburg2014} as given at https://www.cfa.harvard.edu/$\sim$avanderb/k2.html} database by injecting the signal of RR~Cet in the non-variable target light curves based on the analysis of \cite{armstrong2016} from Campaigns 0--4. Then, a shallow signal with an amplitude of $20$\thinspace ppm and frequency of $2.4240$\thinspace c/d was added (to simulate a hypothetical first overtone  component with a characteristic period ratio of $0.746$). Following \cite{kovacs2005}, we used $200$ TFA template light curves selected on a quasi-uniform grid covering the field by the $1100$ brightest members of the field. Exact RR~Cet frequencies were employed to avoid additional complications mentioned in Sect.~2.

\begin{figure}[h]
\centering
\begin{minipage}{0.48\textwidth}
\includegraphics[width=\textwidth]{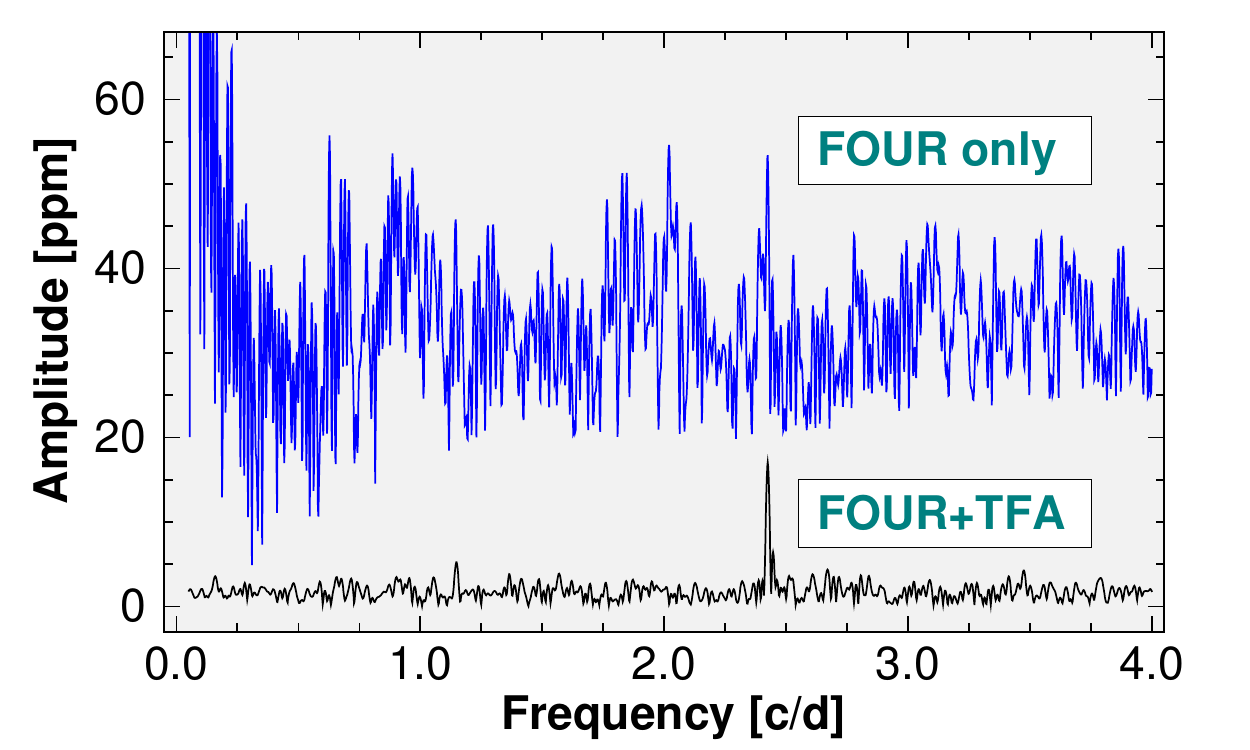}
\caption{Frequency spectra of the residuals obtained after the application of different filtering methods on the large amplitude component of the injected signals in EPIC\,206058354. See text for more details on the signal properties.}
\label{spec-206}
\end{minipage}
\quad
\begin{minipage}{0.48\textwidth}
\includegraphics[width=\textwidth]{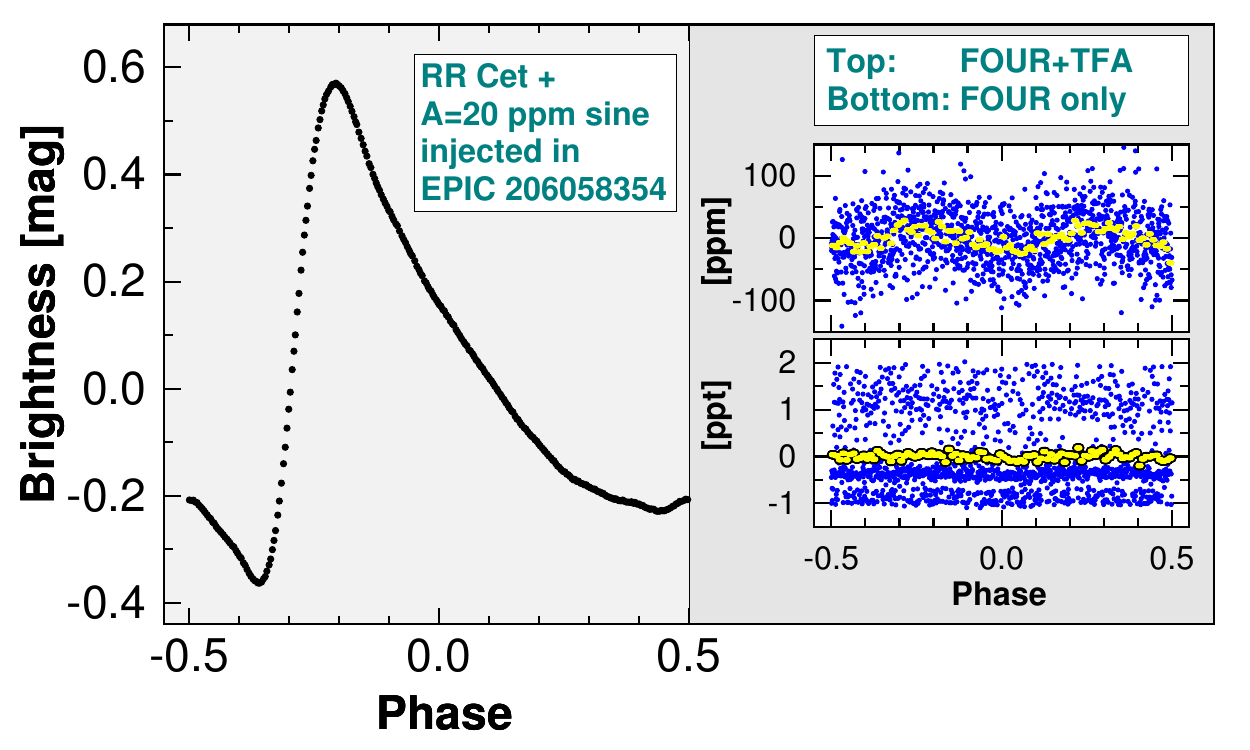}
\caption{Left: full test signal folded by the input period of the large amplitude component. Right: residuals folded by the period of the shallow component, plotted in the bottom panel of Fig.~\ref{spec-206}. The binned values are shown yellow color.}
\label{lc-206}
\end{minipage}
\end{figure}

In general, we got a strong confirmation of the expected signal detection capability of the idea of the full (Fourier$+$TFA) filtering of the data by the large amplitude component prior to the search for additional, small amplitude components. In Figs.~\ref{spec-206} and \ref{lc-206} we show the result of one of our tests performed on EPIC\,206058354, a star exhibiting apparently no variability above the total range of $0.003$\thinspace mag. We see that simple Fourier filtering does not result in a successful detection. Most of the SFF-filtered data contain small trends and signs of variabilities. The high power content at low frequencies is due to these variations. With the application of TFA we can filter out these effects (many of them may have instrumental origins).

%
%

%
%

\section{Conclusions}

In this contribution we indicated the need of using the methodology of data analysis developed in the field of extrasolar planets for searching shallow signals also in other variability studies. The drawback of the systematics filtering is the unavoidable distortion of the signal component that becomes more severe in the signal-dominated regime, whereas it leads only to a relatively small loss of signal power in the systematics-dominated regime. Here the systematics are filtered out more effectively, thereby allowing a considerable gain in signal detection efficiency. The natural extension of the method to signals with partially known signal constituents is briefly described here. In the light of our increasing complex picture on large amplitude pulsators -- e.g., the mystery of the low amplitude modes detected in double-mode pulsators \citep{smolec2017}, the signature of granulation in Cepheids \citep{derekas2017} or the low limit of the Blazhko modulation \citep{benko2015, kovacs2016a} -- we think it is increasingly important to go to the limit allowed by the extremely high quality data supplied by the current and future space observatories.

\acknowledgements{We thank the organizers for their hard work in making this meeting pleasant and memorable. This research has made use of the NASA Exoplanet Archive, which is operated by the California Institute of Technology, under contract with the National Aeronautics and Space Administration under the Exoplanet Exploration Program.}

\bibliographystyle{ptapap}
\bibliography{rrl2017_gk}

\end{document}